\documentclass{revtex4}
\usepackage{color}
\usepackage{graphicx}
\usepackage{dcolumn}
\usepackage{amsfonts}
\usepackage{amsmath}
\usepackage{braket}
\usepackage{amssymb}
\usepackage{bm}
\begin{document}

\title{Exactly solvable tight-binding models on two scale-free networks with identical degree distribution}

\author{Pinchen Xie}
\email{xiepc14@fudan.edu.cn}
\affiliation {Department of Physics, Fudan University,
Shanghai 200433, China}
\affiliation {Shanghai Key Laboratory of Intelligent Information
Processing, Fudan University, Shanghai 200433, China}

\author{Bo Wu}
\affiliation {Department of Applied Mathematics and Statistics, Stony Brook University, Stony Brook, NY 11794, USA}

\author{Zhongzhi Zhang}
\email{zhangzz@fudan.edu.cn}
\homepage{http://www.researcherid.com/rid/G-5522-2011}
\affiliation {Shanghai Key Laboratory of Intelligent Information
Processing, Fudan University, Shanghai 200433, China}
\affiliation {School of Computer Science, Fudan University,
Shanghai 200433, China}

\date{\today}

\begin{abstract}
	We study ideal Bose gas upon two scale-free structures with  identical degree distribution. Energy spectra belonging to tight-binding Hamiltonian are exactly solved and the related spectral dimensions of $\mathcal{G}^1$ and $\mathcal{G}^2$ are obtained  as $d_{s_1}=2$ and $d_{s_2}=2\ln4/\ln3$. We  show Bose-Einstein condensation will only take place upon $\mathcal{G}^2$ instead of $\mathcal{G}^1$.  The topology and thermodynamical property of the two structures are proven to be totally different.
\end{abstract}

\maketitle

\section{Introduction}
{Understanding the spectral structure of Hamiltonians is of key importance in the study of Bose-Einstein condensation (BEC). The underlying dimensionality of a spectrum  determines the occurrence of BEC in most cases. It is well known that for ideal and uniform (homogenous) Bose gases, only the energy spectra related to three or higher dimensions allow  Bose-Einstein phase transition~\cite{Huang1987}. For confined (inhomogenous) Bose gases, it was shown the presence of traps allow BEC to take place in lower dimension notwithstanding~\cite{Bagnato1991, Ketterle1996}. Obviously, the structure and magnitude of present traps have large influence on the spectral structure for quantum gases, and  alter the phase-transition phenomena as well. To further show how the structure of traps influences the thermodynamic behaviors of confined quantum gases, several types of  weakly-coupled discrete traps are investigated over the past decade with the help of network theory ~\cite{Brunelli2004, Buonsante2002, Vidal2011, lyra2014bose, serva2014exactly, deoliveira2009free, DeOliveira2013critical}. For these Bose gases confined by traps with network-like structure, the displayed BEC is topology-induced. For example,  BEC on star-shaped and wheel-shaped network depends on the number of star-arms and wheel spokes~\cite{Brunelli2004, Vidal2011}. The type of Bose-Einstein transition gone through upon diamond hierarchical lattices is also shown to be fully determined by a structural parameter of the lattice-like trap ~\cite{lyra2014bose}. Moreover, a fractal-like energy spectrum is found in Bose gas confined by Apollonian network-shaped traps, which shows self-similarity at the same time~\cite{DeOliveira2013critical}.
As we can see, the topology of different kinds of traps, usually embedded in a $n$-dimensional Euclidean space,  is a key issue to studying  condensation of inhomogenously confined Bose gas. However, still no theory can give a satisfactory explanation how the structures of trap distributions interact with the spectral structure and  determine BEC phenomenon. It is also unclear what  topological invariant can decide the occurrence of BEC or the type of phase transition.  To fill this gap, we pay our attention to traps with scale-free  topology. We will determine whether  the  scale-free characteristics  of weakly coupled traps will govern the BEC.

The  term ``scale-free''   mentioned  before is used to describe a network  with a power-law degree distribution.  In recent decades, dynamics of inhomogenously coupled systems  with  a scale-free topology have been studied extensively since lots of real-world networks were verified to  inherit the same nature  ~\cite{barabasi2009scale, albert2002statistical, pastor-satorras2001epidemic, zhang2009standard, cohen2002percolation, serva2013ising, radicchi2009explosive, herrero2004ising, burioni2001bose, serva2014exactly, deoliveira2009free, duncan2000, cohen2000, leone2002, Chu2010, Wu2013}.  The Hamiltonian defined by  coupling between adjacent vertices is  used to  describe the time evolution of such systems, related to many observable phenomena~\cite{Rammal1984spectrum, DeOliveira2013critical, Mulken2016complex, Maxim2015contact}. The entailed dynamics  are highly dependent on the network topology, as the BEC we focus on in this paper.  Among all the topological features of networks, the  degree distribution  is one of the most important characteristics. Many  networks with a scale-free degree distribution have fractal-like properties, due to the self-similarity underlying  its topological structure~\cite{Song2005, Song2006, barabasi2001deterministic, ravasz2002hierarchical}.  It was also reported that the scale-free characteristics  control the critical phenomena of many physical processes ~\cite{cohen2002percolation, duncan2000, cohen2000, leone2002}. Meanwhile, some literature  suggests that  scale-free characteristics  will not determine phase transitions related to cooperative behaviors and   epidemic spreadings ~\cite{Chu2010, Wu2013}.  Nevertheless  we will show BEC upon weakly coupled traps belongs to the latter.

In this study, we construct two scale-free networks sharing the same degree distribution.  We introduce the tight-binding Hamiltonian to describe weakly coupled traps~\cite{Brunelli2004, Buonsante2002, Vidal2011, lyra2014bose, serva2014exactly, deoliveira2009free, DeOliveira2013critical} . We solve the fractal-like spectra exactly for both networks.   And their spectral dimensions  are analytically calculated~\cite{alexander1982density, rammal1983random}.  The condensed fraction of  low-temperature ideal Bose gas in thermodynamic limit is computed numerically.  And we find these two structures display totally different  BEC phenomena.
}

\section{Construction}\label{construction}
\begin{figure}[h]
\centering
\includegraphics[width=\linewidth]{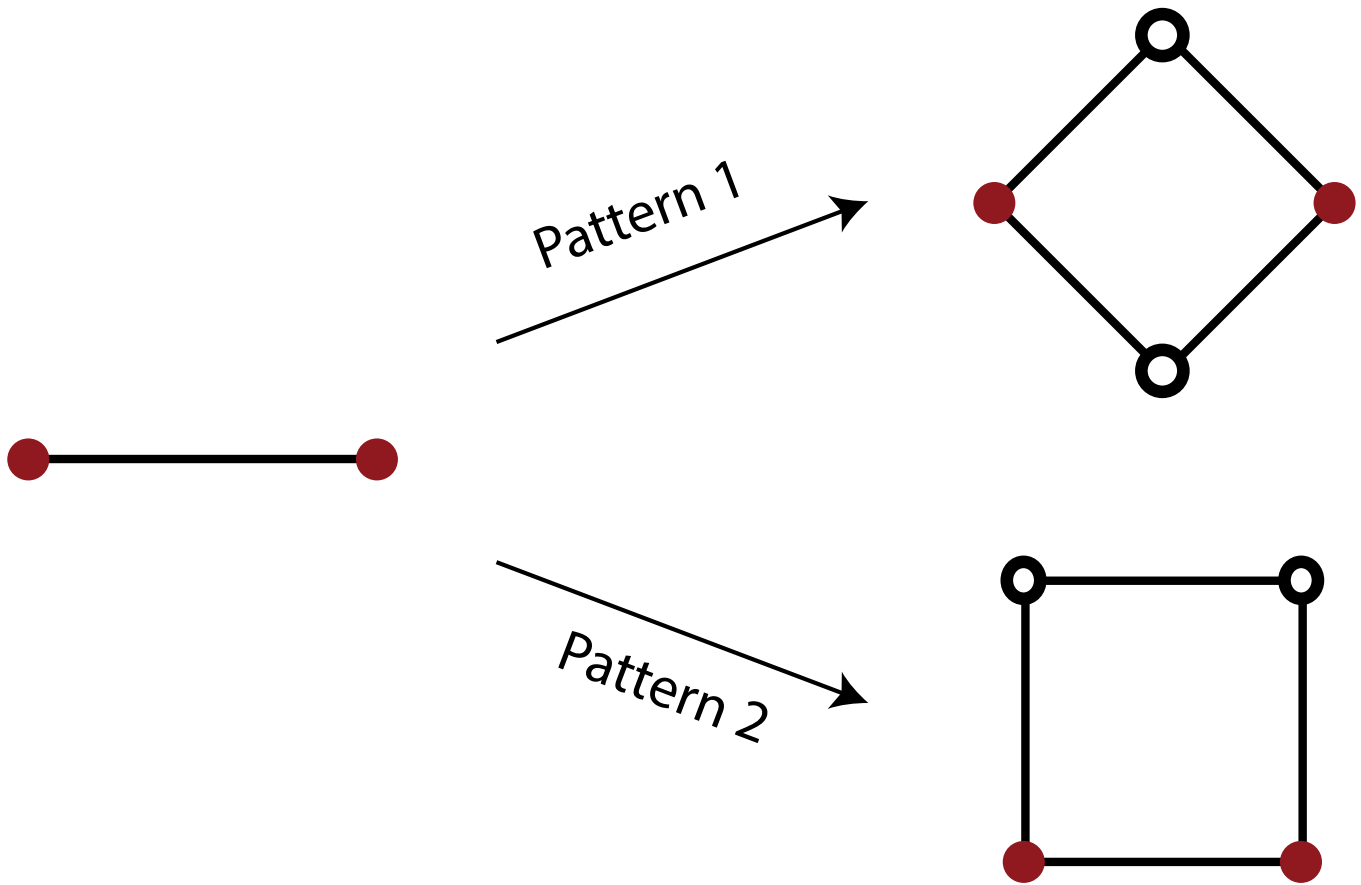}
\caption{(Color online) Two patterns of  construction.}
\label{cons}
\end{figure}

\begin{figure}[h]
\centering
\includegraphics[width=\linewidth]{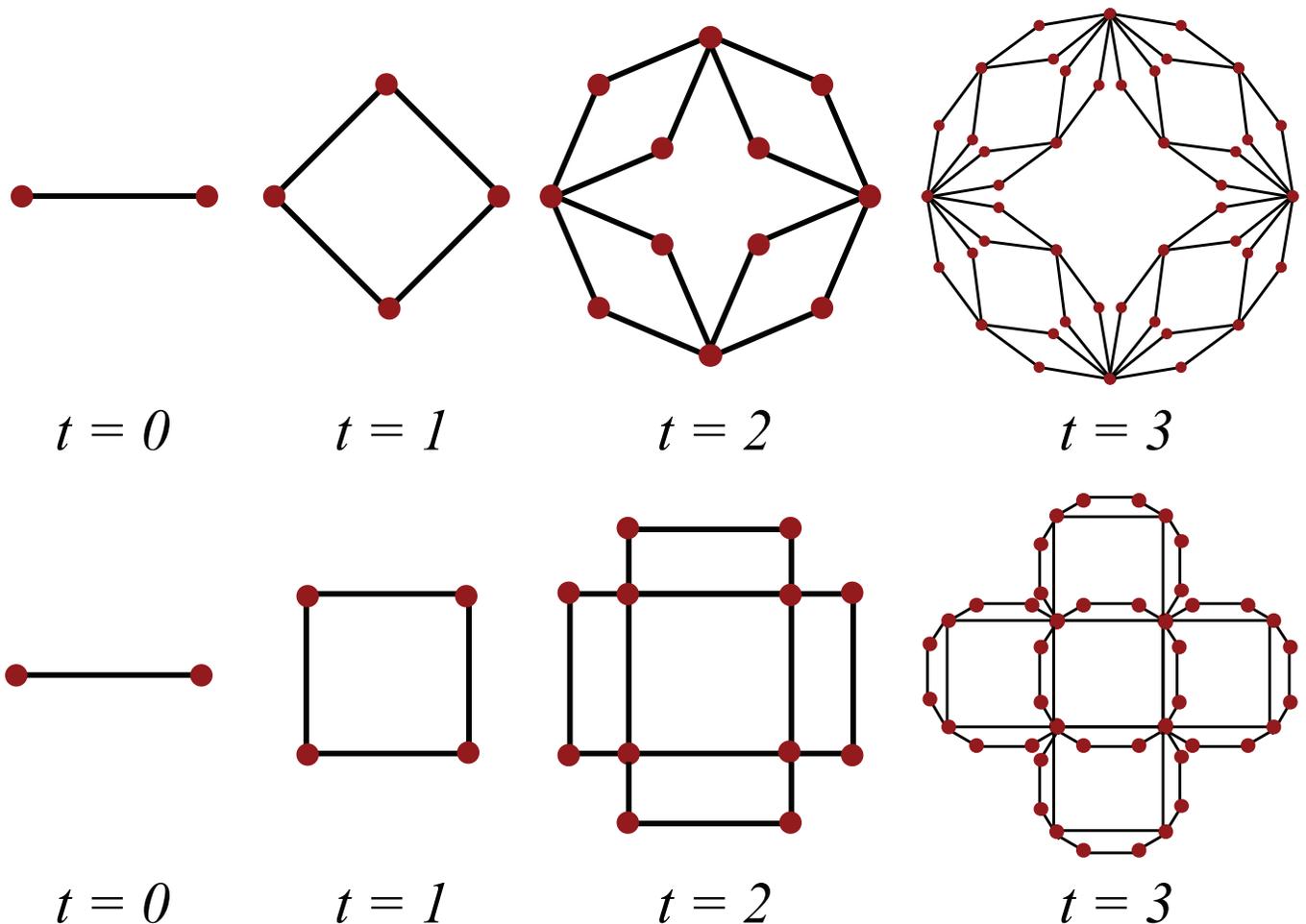}
\caption{(Color online) The first three iterations.}
\label{net}
\end{figure}

	{We construct two scale-free networks  iteratively  in two different patterns} (see Fig.\ref{cons}).   The first few iterations are shown in Fig.\ref{net}. We label them as $\mathcal{G}^1_t$ and $\mathcal{G}^2_t$ respectively according to the pattern {(1 or 2)} and  iterations {($t\in N$)} {they undergo}. $\mathcal{G}^1$ and $\mathcal{G}^2$ denote the networks after infinitely many iterations {($t\rightarrow\infty$)}.

The degree of a vertex is defined as the number of its adjacent vertices. {For both patterns of iteration}, the degree of any vertex is  doubled after one iteration. Thus, $\mathcal{G}^1_t$ and $\mathcal{G}^2_t$  have the same degree distribution $P(k)$, that is the probability distribution of the degrees of vertices over the whole network. After some algebra, we obtained several common characteristics  of the two structures. The  total number of  edges is $E_t=4^t$ for both $\mathcal{G}^1_t$ and $\mathcal{G}^2_t$  and the total number of vertices is $N_t=\frac{2}{3}(4^t+2)$. Their degree distributions {simultaneously obey $P(k)\sim k^{-3}$ when $t\rightarrow\infty $.}

The power-law behavior of $P(k)$ indicates  $\mathcal{G}^1$ and $\mathcal{G}^2$ are both scale-free networks.
{But}  $\mathcal{G}^1$ and $\mathcal{G}^2$ are clustered differently.   Let $L$ denote the typical distance (length of the shortest path) between two randomly chosen nodes of a network. For $\mathcal{G}^2_t$,
\begin{equation}
L\sim \log N_t\sim t
\end{equation}
holds for large $t$.
This means  $\mathcal{G}^2$ is a small-world network~\cite{watts1998collective} with a high clustering coefficient~\cite{zhang2009different}, while $\mathcal{G}^1_t$ is not since its average shortest path length grows faster than $\log N_t$. 

Evidently, the  degree distribution {does not determine all} the global property of a scale-free network. More information is encoded in the pattern the local vertices are organized.

\section{Tight-binding bosons on inhomogeneous structures}\label{tbboson}
\subsection{Tight-binding Hamiltonian}
Label the vertices of any connected graph $\mathcal{G}=(\mathcal{V},\mathcal{E})$ ($\mathcal{V}$ is the set of vertices and $\mathcal{E}$ the set of edges) from $1$ to $N$. {Label}  the edge {connecting} vertex $i$ and $j$ by $e_{ij}$. {Suppose $G$ corresponds to a network of weakly coupled traps, where vertex denotes trap, edge denotes coupling.  Following the tight-binding model defined in~\cite{DeOliveira2013critical}, which is also a simplified LCAO model~\cite{Slater1954}, }the Hamiltonian for non-interacting bosons on $\mathcal{G}$ is
\begin{equation}\label{hamiltonian}
H=\sum_{k\in V}\epsilon_k\ket{k}\bra{k}+\sum_{i,j\in V}h_{ij}\ket{i}\bra{j},
\end{equation}
where $\ket{k}$ is the orbit of  localized boson at vertex $k$ and $\epsilon_k$ the on-site ground energy. {We ignore all  excited states}. $h_{ij}$ is the hopping amplitude between {trap (vertex)} $i$ and $j$. Here we only consider the overlaps of localized states corresponding to adjacent vertices. So $h_{ij}=0$ if vertex $i$ is not connected to vertex $j$. {Suppose the on-site energy is uniform} among all sites,   we  put $\epsilon_k=0$ in Eq.~(\ref{hamiltonian}) without loss of generality.

Next, let us introduce some algebra tools for formulating our model.

Adjacency matrix $A$ is used to describe the connection among vertices of graph $\mathcal{G}$:
\begin{equation}
A_{ij}=
	\begin{cases}
  1		 &\mbox{$e_{ij}\in E$},\\
   0	 &\mbox{$e_{ij}\notin E$}.
   \end{cases}
\end{equation}
The degree matrix $D$ of $\mathcal{G}$ is diagonal, given as $D={\mathrm{diag}}(d_1,d_2,\cdots,d_N)$ where $d_k$ is the degree of vertex $k$.

The transition matrix $M$ of the same graph is defined as $M=D^{-1}A$, which can be normalized as $P=D^{\frac{1}{2}}MD^{-\frac{1}{2}}=D^{-\frac{1}{2}}AD^{-\frac{1}{2}}$. The element of $P$ is thus $P_{ij}=\frac{1}{\sqrt{d_id_j}}$.

{For} homogenous systems such as Bravais lattices, the hopping amplitude is usually taken as  $h_{ij}=\xi A_{ij}$, where $\xi$ is constant. {In the case of scale-free structures}, we assume $h_{ij}=\xi P_{ij}$ to avoid divergence difficulty.

Further, we introduce the {reduced} Hamiltonian $\mathcal{H}=\frac{1}{\xi} H$.  Eq.~(\ref{hamiltonian}) can be rephrased as
\begin{equation}\label{HT}
\mathcal{H}=\sum_{i,j\in V}\frac{1}{\sqrt{d_id_j}}\ket{i}\bra{j}.
\end{equation}

\subsection{Complete energy spectra}
Next, we will solve the  energy spectrum of {$\mathcal{H}$ for $ \mathcal{G}^1$ and $\mathcal{G}^2$} by exact matrix renormalization. Similar derivation of spectrum for other renormalizable structures can be found at ~\cite{Huang2015, Xie2015}.

Let $D_{i,t}$ and $A_{i,t}$ denote the degree matrix and adjacency matrix of $\mathcal{G}^i_t (i=1,2)$.

For $i=1$, {by proper permutation of rows and columns,   $D_{i,t}$ and $A_{i,t}$ write}
\begin{equation}
D_{1,t}=\left(
\begin{array}{cccc}
2I & 0 \\
0 & 2D_{1,t-1}
\end{array}
\right),
\end{equation}
\begin{equation}
A_{1,t}=\left(
\begin{array}{cccc}
0 & J^T \\
J & 0
\end{array}
\right)
\end{equation}
where $I$ is the identity matrix of order {$\Delta_t=N_t-N_{t-1}$. And $N_{t-1}\times \Delta_t $ matrix $J$ represents the adjacency  among new vertices and the old ones.}
The hopping matrix (i.e. the normalized transition matrix of $\mathcal{G}^1_t$) is
\begin{equation}
P_{1,t}=D_{1,t}^{-\frac{1}{2}}A_{1,t}D_{1,t}^{-\frac{1}{2}}=\left(
\begin{array}{cccc}
0 & \frac{1}{2}J^TD_{1,t-1}^{-\frac{1}{2}} \\
\frac{1}{2}D_{1,t-1}^{-\frac{1}{2}}J & 0
\end{array}
\right).
\end{equation}
The characteristic polynomial of $P^1_t$ is thus
\begin{equation}\label{char}
\det(\lambda-P_{1,t})=\lambda^{N_{t}-N_{t-1}}\det\left(\lambda-\frac{1}{4\lambda}D_{1,t-1}^{-\frac{1}{2}}JJ^TD_{1,t-1}^{-\frac{1}{2}}\right),
\end{equation}
using the identity
\begin{equation}
  \det\left(\begin{array}{cccc}
A & B \\
C & D
\end{array}\right)=\det A\cdot \det(D-CA^{-1}B).
\end{equation}

{Let $j_{mn}$ denote the $(m,n)$-entry of $J$. The symmetric matrix $JJ^{T}$ is represented by
\begin{equation}
	(JJ^{T})_{mn}=\sum_{l} j_{ml}j_{nl}.
\end{equation}
For $m=n$, $(JJ^{T})_{mn}$ is exactly the degree of node $m$ in $\mathcal{G}^1_t$, i.e., twice its degree in  $\mathcal{G}^1_{t-1}$. For $m\neq n$, $(JJ^{T})_{mn}$ is $2$ if node $m$ and node $n$ is previously adjacent in  $\mathcal{G}^1_{t-1}$, otherwise $0$.

Hence one obtains
}

\begin{equation}
  JJ^T=2D_{1,t-1}+2A_{1,t-1},
\end{equation}
 which leads to
\begin{equation}\label{rec}
\det(\lambda-P_{1,t})={\tfrac{1}{2}}^{N_{t-1}}\lambda^{N_{t}-2N_{t-1}}\det((2\lambda^2-1)-P_{1,t-1}).
\end{equation}

The recursive relation Eq.~(\ref{rec}) indicates, if $\lambda$ is an eigenvalue of $P_{1,t}$, then  $R_1(\lambda)=(2\lambda^2-1)$ is  an eigenvalue of $P_{1,t-1}$ with the same degeneracy unless $\lambda=0$. Conversely, for any eigenvalue $\eta$ of $P_{1,t-1}$, the {inverse of $R_1$} gives its two descendents with the same degeneracy as $\eta$,  {unless $\eta=-1$ (notice $R_1(-1)=0$).  Hence the degeneracy of  the exceptional eigenvalue $0$  should be  $\frac{N_t}{2}$ to ensure the spectrum of $P_{1,t-1}$ is complete.}

{Subsequently}, the spectrum of $P_{1,t}$, denoted by   $\sigma_{1,t}$,  can be analytically determined from the initial spectrum  $\sigma_{1,1}=\{\cos0, \cos\frac{\pi}{2}, \cos\frac{\pi}{2}, \cos\pi\}$, given as
\begin{equation}
\sigma_{1,t}=\{E_i\}
=\bigcup_{\mbox{\tiny
$\begin{array}{c}
0\leqslant k \leqslant 2^r\\
0\leqslant r\leqslant t
\end{array}$}}\bigg{\{} \cos\frac{k\pi}{2^r}\bigg{\}}.
\end{equation}

The symmetry of  $\sigma_{1,t}$ with respect to $0$ is obvious  and the lowest energy is always $E_0=-1$.

The integrated  density of states (IDOS) $f(\varepsilon )$ is defined to be  the number of states between $E_0$ and $E_0+\varepsilon$ {($0\leq \varepsilon\leq 2$)} divided by the total number of states. 

For homogenous systems such as a single particle in a 3-D cavity, it is well known that $f(\varepsilon)\propto \varepsilon^{\frac{3}{2}}$. {For inhomogenous systems also showing power-law behaviors  $f(\varepsilon)\propto \varepsilon^{d_s/2}$ near the band bottom($\varepsilon\ll 1$), the exponent $d_s/2$ can be  irrational. $d_s$ is said to be the spectral dimension of a  structure related to certain Hamiltonian~\cite{alexander1982density, rammal1983random}. It was reported that spectral dimension is a crucial index categorizing  the universality classes of topology-induced Bose-Einstein  transitions~\cite{lyra2014bose}. Also, it has  been  proven that the phase transition breaking a continuous symmetry can not take place on systems with a spectral dimension not greater than 2~\cite{burioni1996universal, burioni2005random, cassi1992phase, cassi1993spectral}. The theoretical determination of spectral dimension is not an easy task. However,
by appropriate renormalization, analytical results on spectral dimension are still obtainable for several fractal-like structures~\cite{Burioni1999}. 
}

\begin{figure}
\centering
\includegraphics[width=0.4\textwidth]{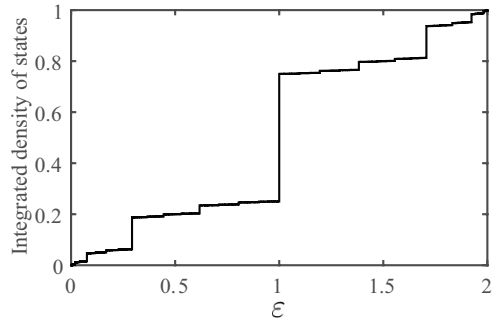}
\caption{The fractal-like integrated density of states  for $\mathcal{G}^1_{12}$.  }
\label{I_idos}
\end{figure}

To compute the spectral dimension of our networks, we first pay attention to the  IDOS  $f(\varepsilon )$  corresponding to $\mathcal{G}^1$. Transforming  $R_1$ for expressing the iterative relation {for}  the relative energy $\varepsilon$,  one obtains $\tilde{R}_1(\varepsilon)=2(\varepsilon-1 )^2$. The following relation holds for small $\varepsilon$ because $\mathcal{G}^1$ is invariant under iteration :
\begin{equation}\label{cdos}
\begin{split}
	f(\varepsilon )&=(1-f(\tilde{R}_1(\varepsilon))) \lim_{t\rightarrow\infty}\tfrac{N_{t-1}}{N_t}\\
	&=\frac{1}{4}f(2-\tilde{R}_1(\varepsilon))\\
	&\approx\frac{1}{4}f(4\varepsilon).
	\end{split}
\end{equation}
Hence $f(\varepsilon )\propto \varepsilon^1$ near the origin.
 The spectral dimension of $\mathcal{G}^1$ is thus $d_{s_1}=2$. Fig.~\ref{I_idos} gives  schematic representation of the spectrum.

\begin{figure}[h]
\centering
\includegraphics[width=0.4\textwidth]{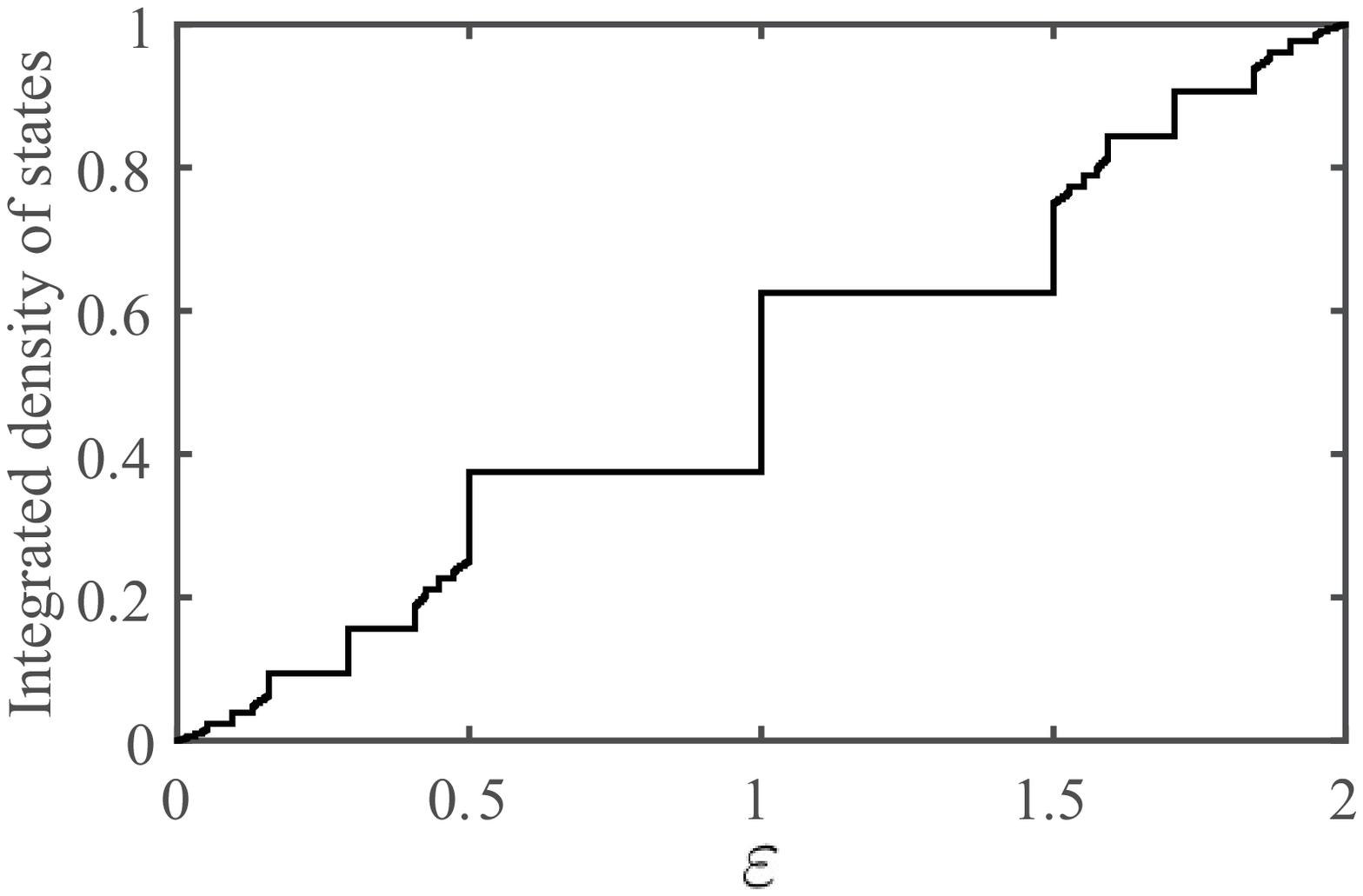}
\caption{The  integrated density of states  for $\mathcal{G}^2_{12}$.}
\label{II_idos}
\end{figure}

By the same method, we calculate the spectrum for  $\mathcal{G}^2_t$ and obtain the iterative expression for the characteristic polynomial:
\begin{equation}\label{char2}
\begin{split}
\det(\lambda-P_{2,t})=&2^{-N_{t-1}}\cdot(\lambda^2-\tfrac{1}{4})^{\frac{N_t-3N_{t-1}}{2}}\\
&\cdot \lambda^{2N_{t-1}}\cdot \det(\tfrac{2\lambda^2-1}{\lambda}-P_{2,t-1}).
\end{split}
\end{equation}
Now the iteration is  formulated as  $R_2(\lambda)=\frac{2\lambda^2-1}{\lambda}$. By taking the inverse of $R_2$, we are able to generate the spectrum of any order from the initial one.
The exceptional eigenvalues here are  $\frac{1}{2}$,  $-\frac{1}{2}$  and $0$. Their degeneracies are $(N_t-3N_{t-1}+2)/2$, $(N_t-3N_{t-1}+2)/2$ and $N_{t-1}$ respectively according to Eq.~(\ref{char2}).
Since $\sigma_{2,1}=\{1,0,0,-1\}$, the energy spectrum $\sigma_{2,t}$ for $\mathcal{G}^2_t$ is symmetric with respect to zero and the lowest-energy  is still $E_0=-1$.
For $\mathcal{G}^2$, when  $\varepsilon$ is small, the invariance of $f(\varepsilon)$ requires
\begin{equation}
\begin{split}
	f(\varepsilon)&=f(2\varepsilon-1-\tfrac{1}{\varepsilon-1})\lim_{t\rightarrow\infty}\tfrac{N_{t-1}}{N_t}\\
	&\approx \frac{1}{4}f(3\varepsilon).
\end{split}
\end{equation}
Thus
\begin{equation}
f(\varepsilon)\propto \varepsilon^{\frac{d_{s_2}}{2}}
\end{equation}
near the band bottom and the spectral dimension is
\begin{equation}
  d_{s_2}=2\frac{\ln4}{\ln3}=2.524.
  \end{equation}

The  IDOS for finite-size network is shown in Fig.~\ref{II_idos}.

By so far, the spectra related to the two networks have shown unique fractal-like structures, which will leads to different behaviors of tight-binding particles.

\section{Different cryogenic behaviors of Bose gas}\label{DCBB}
 In this section we investigate the cryogenic behaviors of non-interacting Bose gas on $\mathcal{G}^1$ and $\mathcal{G}^2$  and check whether Bose-Einstein condensation will take place.

Suppose there are  $N_p$ bosons on the structures. The particle density is defined as $\gamma=\frac{N_p}{N_t}$. {To approach the   thermodynamic limit, we fix $\gamma$ and let $t \rightarrow \infty$.}
Bose-Einstein statistics gives the expected number of bosons in  state $i$:
\begin{equation}
	n_i=\frac{1}{z^{-1}e^{\beta\xi \varepsilon}-1}
\end{equation}
where $\beta=\frac{1}{k_BT}$ and the fugacity $z=e^{\beta(\mu-\xi E_0)}$. $\mu$ is the chemical potential.
\begin{equation}\label{condensed}
	n_0=\frac{1}{z^{-1}-1}
\end{equation}
 is the number of condensed particles and $f=\frac{n_0}{\gamma N_t}$ the condensed fraction.

The normalization condition requires
\begin{equation}\label{NL}
	\sum_i n_i=\gamma N_t.
\end{equation}
Transforming  Eq.~(\ref{NL}) into integral, one obtains
\begin{equation}\label{sum}
\int_0^2  \frac{\rho(\varepsilon)}{z^{-1}e^{\beta \xi \varepsilon}-1} d\varepsilon =\gamma
\end{equation}
where $\rho(\cdot)$ is the states density  and $\varepsilon$ the relative energy.

Let $\tilde{T}=\frac{k_BT}{\xi}$ be the dimensionless temperature. Substitute $\xi \beta$ with $\tilde{\beta}=1/\tilde{T}$.
 Eq.~(\ref{sum}) is rewritten as
\begin{equation}
\int_0^2  \frac{\rho(\varepsilon)}{z^{-1}e^{\tilde{\beta} \varepsilon}-1} d\varepsilon =\gamma.
\end{equation}

Below the critical temperature where the Bose-Einstein phase-transition occurs, $z$ is always $1$.
\begin{figure}[t]
\centering
\includegraphics[width=\linewidth]{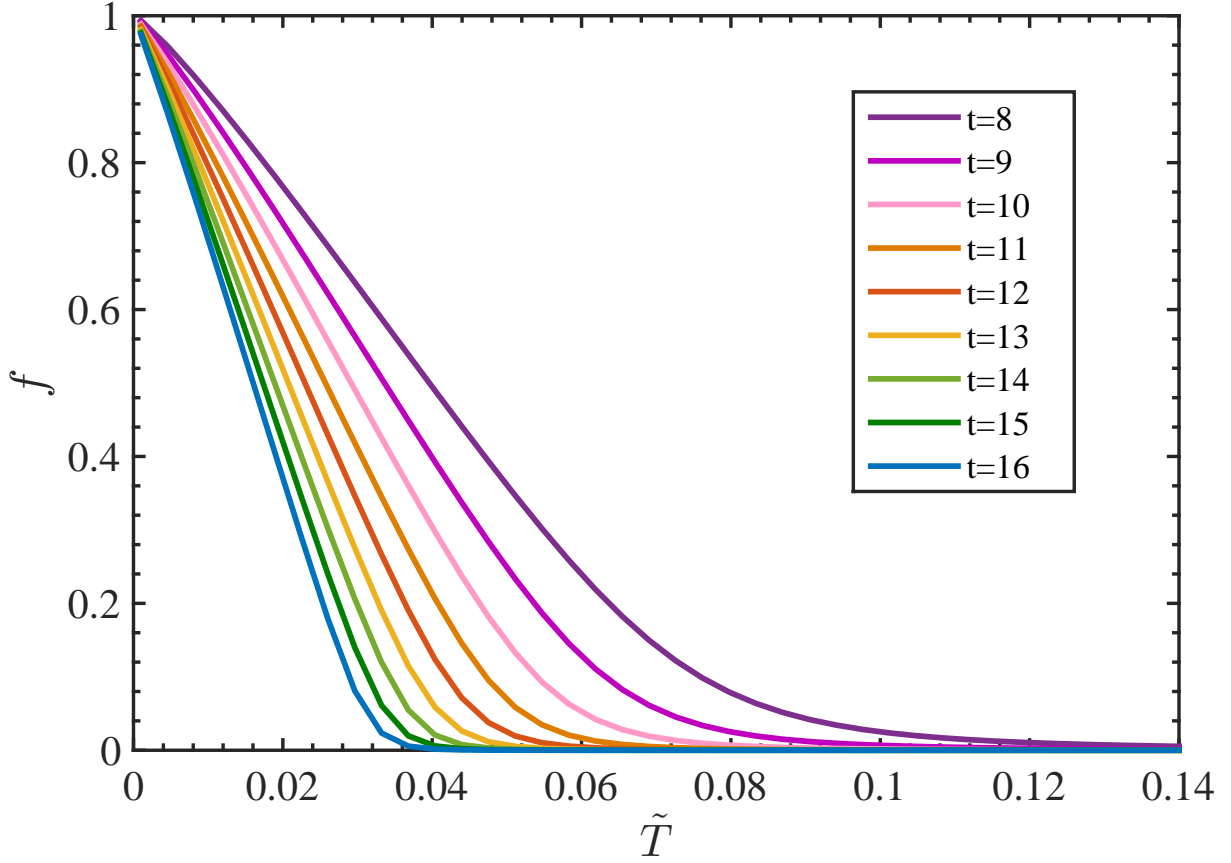}
\caption{(Color online) Condensed fraction  $f$  as  a function of $\tilde{T}$ for $G_t^1$ ($\gamma=0.2$). The finite-size effect  {decreases as $t$ increases}. }
\label{I_condensed}
\end{figure}
In Fig.~\ref{I_condensed} we present the relation between the condensed fraction and the dimensionless temperature with finite-size effects. There is no sign of  first order phase-transition, which is proved  analytically as follows.

Suppose  $\mathcal{G}^1$  allows the occurrence of BEC transition in thermodynamic limit. The uncondensed fraction of bosons at the (dimensionless) critical temperature $T_{c_1}$ is
\begin{equation}
f_u=\frac{1}{\gamma}\lim_{\epsilon\to 0}\int_\epsilon^2  \frac{\rho(\varepsilon)}{e^{\beta_{c_1}\varepsilon}-1} d\varepsilon >\frac{1}{\gamma}\int_{\epsilon_1}^{\epsilon_2}  \frac{\rho(\varepsilon)}{e^{\beta_{c_1}\varepsilon}-1} d\varepsilon
\end{equation}
 where $\beta_{c_1}=\frac{1}{T_{c_1}}$, $\epsilon_1< \epsilon_2 \ll 1$.

 By the approximation $\rho(\varepsilon)\approx k \frac{d (\epsilon ^{\frac{d_{s_1}}{2}})}{d\epsilon}$, the last integral becomes
 \begin{equation}
 \begin{split}
 		\frac{1}{\gamma}  \int_{\epsilon_1}^{\epsilon_2}  \frac{\rho(\varepsilon)}{e^{\beta_{c_1}\varepsilon}-1} d\varepsilon
 &\approx\frac{1}{\gamma}  \int_{\epsilon_1}^{\epsilon_2}  k \frac{d (\epsilon ^{\frac{d_{s_1}}{2}})}{d\epsilon} \cdot \frac{1}{e^{\beta_{c_1}\varepsilon}-1} d\varepsilon \\
 &=\frac{k}{\gamma}\int_{\epsilon_1}^{\epsilon_2}  \frac{1}{e^{\beta_{c_1}\varepsilon}-1} d\varepsilon
 \end{split}
\end{equation}
 where $k$ is a positive constant.

{Since} $\epsilon_1$ and $\epsilon_2$ are  small, $e^{\beta_{c_1}\epsilon}\approx 1+\beta_{c_1}\epsilon$ holds for $\epsilon\in [\epsilon_1,\epsilon_2] $. It follows that
 \begin{equation}\label{ineq}
\frac{k}{\gamma}\int_{\epsilon_1}^{\epsilon_2}  \frac{1}{e^{\beta_{c_1}\varepsilon}-1} d\varepsilon \approx \frac{k}{\gamma \beta_{c_1}}\ln\left(\frac{\epsilon_2}{\epsilon_1}\right).
\end{equation}

Thus,
\begin{equation}\label{cont}
	1\geq f_u>\frac{k}{\gamma\beta_{c_1}}\ln\left(\frac{\epsilon_2}{\epsilon_1}\right).
\end{equation}

  $\epsilon_2 /\epsilon_1$ has no finite upper bound. {This contradicts} Eq.~(\ref{cont}) when $T_{c_1}>0$.   {Hence}  positive $T_{c_1}$ does not exist for $\mathcal{G}^1$. {In consequence,} Bose-Einstein condensation won't take place in low temperature.

\begin{figure}[t]
\centering
\includegraphics[width=\linewidth]{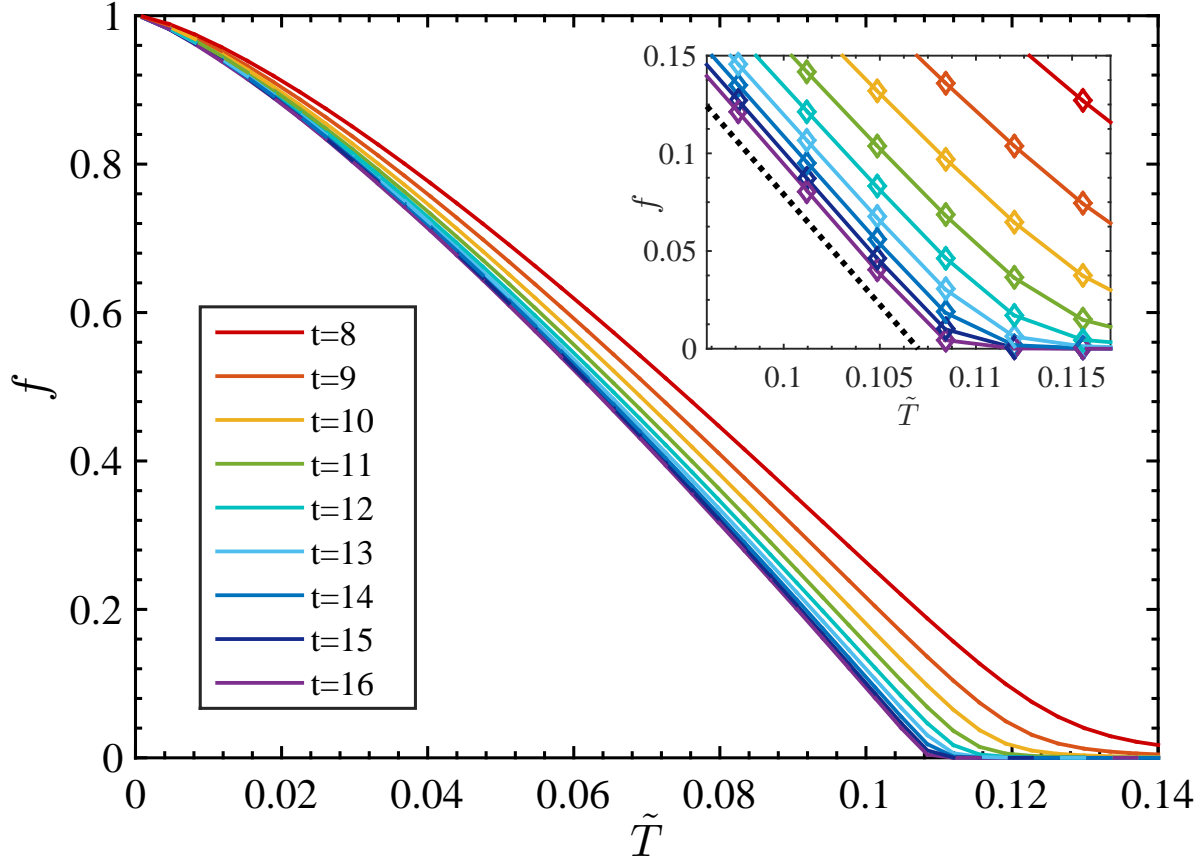}
\caption{(Color online) Condensed fraction  $f$  as  a function of $\tilde{T}$ for $\mathcal{G}_t^2$ ($\gamma=0.2$). The dotted line corresponds to the critical behavior. }
\label{II_condensed}
\end{figure}

Will the massive bosons behave differently on $\mathcal{G}^2$ in low temperature?
A schematic representation of the relation between condensed fraction for network $\mathcal{G}^2$  and $\tilde{T}$ is given in Fig. \ref{II_condensed}. The curves converge quickly, suggesting the transition temperature when thermodynamic limit is approached.  For $\gamma=0.2$, we obtain the transition temperature  $\tilde{T}_c\approx 0.107$ and  $f\propto|T-T_c|^1$ near the critical point.

By further numerical computation, the dependence of the critical temperature on $\gamma$ and a series of  critical exponents can be obtained, which are not the main interest of this study. Related discussions on these can be found at ~\cite{lyra2014bose, hall1975scaling}. We  claim that the BEC in $\mathcal{G}^2$ belongs to the universality class of the ideal BEC in networks with spectral dimension $d_s\approx 2.524$.

Moreover, the state $\psi_c$ of the  condensate can be determined analytically. Since hopping amplitude is negative ($\xi<0$),  $\psi_c$ is related to  the eigenvector of  $P_{2,t}$ with respect to the eigenvalue $1$, which is usually called the equilibrium distribution or steady state of random walks depicted by the transition matrix. With finite-size effects ($t$ is finite), $\psi_c$ is expressed as
\begin{equation}\label{LCAO}
	\psi_c=A \sum_i \frac{d_i}{N_t} \ket{i}.
\end{equation}
A is the normalization constant. The sum is taken over all the vertices.

Eq.~(\ref{LCAO}) indicates the state of the condensate  follows the degree distribution though the occurrence of phase transition is determined by more in-depth topology. Not only for $\mathcal{G}^2$, this consequence holds true  for   all discrete structures with a tight-binding Hamiltonian. If the structure is a random regular graph, the lowest state is simply the unweighted combination of all tight-binding local orbits {up to  phase difference}.

\section{Conclusion}

Tight-binding models upon two scale-free networks  with identical degree distribution $P(k)\sim k^{-3}$ were investigated.

By renormalization, we iteratively obtained the {fractal-like}  spectra of the two networks and determined their spectral dimensions ($d_{s_1}=2$, $d_{s_2}=2 \ln4/\ln3$). Suggested by the value of $d_{s_1}$, we  analytically proved BEC would not take place in $\mathcal{G}^1$. On the contrary, with the same scale-free  degree distribution, the structure of $\mathcal{G}^2$ allows the occurrence  of the Bose-Einstein phase-transition. Meanwhile, the BEC in $\mathcal{G}^2$ belongs to the universality class of the ideal BEC, related to  spectral dimension $d_s=2 \ln4/\ln3$.
  We also found the state $\psi_c$ for the condensate, which was determined by the degree distribution of the structure.

  The divergent behaviors  of the two structures  give a good example how the topology as well as thermodynamical property of networks varies regardless of {scale-free characteristics.} The divergence not merely lies in  several critical exponents but the  occurrence of phase-transition. The scale-free characteristics  do not always play a important role in  dynamical systems governed by equations of the form $dx_i/dt=\sum k_{ij}(x_j-x_i)$, related to diffusion, relaxation, etc.

\acknowledgments
 This work was supported by the National Natural Science Foundation of China under Grant No. 11275049.
 
\bibliographystyle{unsrt}
\bibliography{reference}

\end{document}